\documentclass[twocolumn,pra,superscriptaddress,showpacs,floatfix]{revtex4}
\usepackage{epsfig}
\usepackage{times}
\usepackage{amsmath,color}

\begin{document}

\title{Phase separation of trapped spin-imbalanced Fermi gases in one-dimensional optical lattices}

\author{F. Heidrich-Meisner}
\affiliation{Department of Physics, Arnold Sommerfeld Center for Theoretical Physics,
and Center for NanoScience, Ludwig-Maximilians-Universit\"at M\"unchen, D-80333 M\"unchen, Germany}
\author{G. Orso}
\affiliation{Department of Physics, Arnold Sommerfeld Center for Theoretical Physics,
and Center for NanoScience, Ludwig-Maximilians-Universit\"at M\"unchen, D-80333 M\"unchen, Germany}
\author{A.E. Feiguin}
\affiliation{Department of Physics and Astronomy, University of Wyoming, Laramie, WY 82071, USA}

\date{May 5, 2010}

\begin{abstract}
We calculate the density profiles of a trapped spin-imbalanced Fermi gas with attractive interactions in a one-dimensional optical lattice, using both the local density approximation (LDA) and density matrix renormalization group (DMRG) simulations. Based on the exact equation of state 
obtained by Bethe ansatz,  LDA predicts that the gas phase separates into  shells with a partially polarized core and fully paired wings, where the latter occurs below a critical spin polarization.
This behavior is also seen in  numerically exact DMRG calculations at sufficiently large particle numbers.
Unlike the continuum case, we show that the critical polarization is a non monotonic function of the interaction strength and vanishes in the limit of large interactions.
\end{abstract}

\maketitle

\section{Introduction}

 The physics of population imbalanced  Fermi gases in one dimension 
is currently attracting  substantial interest 
since in these systems, the one-dimensional (1D) counterpart of the Fulde-Ferrell-Larkin-Ovchinnikov  (FFLO) state \cite{fulde64,larkin64} can be realized. This inhomogeneous superfluid 
state of fermions was suggested long time ago, yet its unambiguous observation  in condensed matter systems has turned out to be quite challenging \cite{budzin05,radovan03,bianchi04,ronning05}.
Ultracold atomic gases offer a perfect avenue to search for exotic pairing states  such as the 
 FFLO state, due to the clean experimental conditions and the tunability of interactions and imbalance \cite{bloch08,giorgini08,ketterle08}.
First experiments for three-dimensional (3D) systems at by Ketterle and co-workers \cite{zwierlein06a,shin08} and by Hulet and co-workers \cite{partridge06,partridge06a} as well as theoretical work for 3D \cite{sheehy07} 
indicate that the chances of observing the FFLO state in 3D are slim since this phase 
appears to be stable only in a narrow window of the BCS-BEC crossover phase diagram.

A promising direction to detect the FFLO state is to confine atoms in 1D configurations. In this limit the  Fermi surface reduces to two points, $k_{F\uparrow}$ and $k_{F\downarrow}$, 
forcing all Cooper pairs to share the same wave-vector $Q_{\mathrm{FFLO}}=k_{F,\uparrow}-k_{F,\downarrow}$.
For systems of fermions with attractive interactions -- usually described by the Hubbard model or the associated continuum model -- it has indeed  been shown
by means of  analytical and numerical techniques that  any finite imbalance drives the system into the 1D FFLO state, which  has algebraically decaying pair-pair correlations in the s-wave channel, 
modulated with the FFLO wavevector $Q_{\mathrm{FFLO}}$ \cite{yang01,hu07,orso07,feiguin07,tezuka08,batrouni08,rizzi08,casula08,luescher08,zhao08}.

In the more general case of a two-channel model that also accounts for the formation of composite molecules, the FFLO  phase still emerges on the BCS side in the imbalanced case, yet eventually gives room to a Bose-Fermi mixture 
phase  \cite{baur09,hm09c}. The FFLO phase also exists in mass-imbalanced systems in 1D \cite{batrouni09,wang09,burovski09,orso09,lu09}, 
which could be realized in Li-K mixtures (see Ref.~\cite{taglieber08}). It should further be possible to drive the  FFLO state in a globally balanced mixture by using
spin-dependent potentials \cite{orso08} or spin-dependent optical lattices \cite{feiguin09b}.
Furthermore, FFLO physics has  been studied in 3D arrays of  coupled 1D systems \cite{parish07,luescher08,zhao08}, the two-leg ladder geometry \cite{feiguin09},
and the possibility of inhomogeneous pairing states in multi-band systems has been explored in Ref.~\cite{zhang09}.  
Finally, we mention  that the spectral function of attractively interacting fermions with  spin-imbalance has been  calculated and discussed in Ref.~\cite{feiguin09a}.
 
Experimentally,  the simplest observables that can be measured in trapped Fermi gases are the density  profiles \cite{zwierlein06a,shin08,partridge06,partridge06a,nascimbene09} of the two spin components  \cite{zwierlein06a,shin08,partridge06,partridge06a,nascimbene09}. On the basis of the exact
Bethe ansatz solution for the homogeneous system, it has been independently predicted   by Hu {\it et al.} \cite{hu07} and one of the authors \cite{orso07} that the trapped gas phase-separates into two shells:
a partially polarized core, and either fully paired or fully polarized wings.  
Very recently, this theoretical prediction has been addressed and verified in an experiment with ultracold fermions in 1D wave-guides at Rice \cite{liao09}. 
 
In the case of an optical lattice, the shell structure
of spin-imbalanced fermions  was studied with the density matrix renormalization group (DMRG) method in  Refs.~\cite{feiguin07,tezuka08,machida08}
and with quantum Monte Carlo (QMC) in Ref.~\cite{batrouni08}.  
In Ref.~\cite{feiguin07} by some of us, clear evidence for the presence of FFLO correlations of the 
partially polarized core has been reported on systems with up to 80 fermions. In another DMRG work by Tezuka and Ueda \cite{tezuka08}, a
phase diagram for the shell structure has been proposed, with lower critical polarizations for the emergence of the fully polarized wings.
In Ref.~\cite{tezuka08}, the presence of the fully polarized wings was deduced from the presence of certain local maxima in the density difference profile in the outer regions of the trap.
However, neither of these studies observed the fully paired wings,
which have been seen in QMC calculations, in the continuum case \cite{casula08}.
 
\begin{figure}[t]
\epsfig {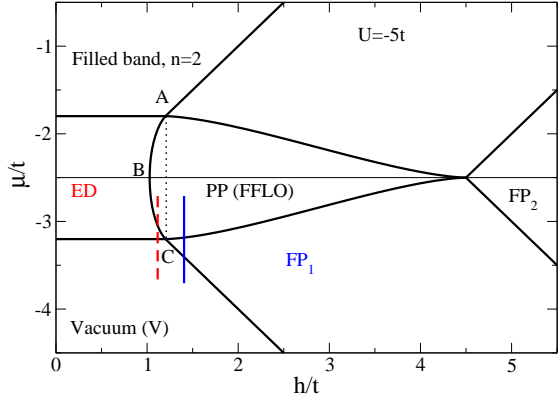}
\caption{
(color online)
Grand-canonical phase diagram of the 1D attractive Hubbard model \eqref{one} for $U=-5t$ (compare \cite{essler-book}). The vertical lines are LDA trajectories  yielding different shell structures for the trapped gas:  the dashed (red)  and the solid (blue) lines correspond to configurations with fully paired (ED) and fully polarized (FP$_1$)  wings, respectively.
The dotted line  corresponds to a special value of the magnetic field, $h=\sqrt{U^2+16t^2}/2-2t$, where the partially polarized phase extends to the edge of the cloud.
}
\label{fig0}
\end{figure}

The purpose of this work is to clarify this open point, namely the presence of fully paired wings in an optical lattice. As a result,  numerically exact DMRG simulations indeed show the existence of fully paired wings at  low spin polarizations. 
To that end we perform large scale DMRG \cite{white92b,schollwoeck05} calculations of the 1D Hubbard model with (effectively) up to 320 fermions in a harmonic trap, which is comparable to the particle numbers typically encountered in experiments \cite{bloch08,moritz05}. We carefully compare the DMRG results with those obtained  starting from the exact solution
of the homogeneous lattice gas,  including the effect of  the trap via the local 
density approximation (LDA), as previously done for the continuum case \cite{orso07,hu07}. 

Our study unveils that the fully paired wings are  present below a critical spin  polarization, which 
 we calculate as a function of effective density and interaction strength. In contrast with
the continuum case, in the strong coupling regime, the fully paired wings shrink in size
as the attraction strength increases. As we shall see, this effect is due to the enhanced inertia of the
molecules with respect to unpaired fermions.
 
The outline of this work is as follows. In Sec.~\ref{sec:model}, we shall
introduce the model and discuss in detail how we obtain the density profiles of the trapped gas,  
using either DMRG simulations or  LDA applied to the exact equation of state for the homogeneous system. Our main results are contained in Sec.~\ref{sec:results}, where we compare the two methods  and discuss density profiles as well as the critical polarization.

\section{Model, methods and set-up}
\label{sec:model}
\subsection{Hamiltonian}

Our simulations are performed for  the 1D Hubbard model with an attractive onsite interaction $U<0$: 
\begin{equation}
H_0 =   -  t \sum\limits^{L-1}_{i=1, {\sigma}} \left(c^\dagger_{i\sigma} c_{i+1\sigma}+h.c.\right)
+ U \sum\limits^L_{i=1} n_{i\uparrow} n_{i\downarrow}
\label{one}
\end{equation}
where $c^\dagger_{\ell\sigma}$ creates a fermion with spin $\sigma=\uparrow,\downarrow$  at 
site $l$, $n_{\ell\sigma}=c^\dagger_{\ell\sigma}c_{\ell \sigma}$, $n_{\ell}=n_{\ell\uparrow}+n_{\ell\downarrow}$ is the local density, 
and $t$ is the hopping matrix element. 
We define $x=ia$, where $a$ is the lattice spacing, set to unity. 
The local polarization is defined as $\langle s_i\rangle = \langle n_{i,\uparrow}-n_{i\downarrow}\rangle$,  the filling factor is $n=N/L$ where $N_{\sigma}=\sum_{i=1}^L \langle n_{i\sigma} \rangle$ and $N=\sum_{\sigma} N_{\sigma}$. Analogously, we define the densities for 
spin up and down as $n_{\sigma}=N_{\sigma}/L$.

Our main interest will be in the case of trapped fermions for which we can write the Hamiltonian
as 
\begin{eqnarray}
H &=& H_0+H_{\mathrm{trap}} \\
H_{\mathrm{trap}}& = & V\sum\limits^L_{i=1}(x-L/2)^2n_i \label{eq:strap}\,, 
\end{eqnarray}
where in the second line, $V$ is the trapping potential.
To characterize the trapped system, a suitable choice of variable is the total
polarization   $p=(N_{\uparrow}-N_{\downarrow})/N$ and the effective density $\rho_{\mathrm{eff}}=N\sqrt{V}$.
In the following we fix the energy scale by setting $t=1$.

\subsection{Bethe ansatz and Local Density Approximation} 

The density profiles of the two component Fermi gas can be obtained starting from the exact Bethe ansatz solution of the homogeneous system (\ref{one}),  taking into account the effect of the trap via the local density approximation. 

\subsubsection{Homogeneous system} 

The exact ground state energy per site $E(n_\uparrow,n_\downarrow)$ of the gas
in the thermodynamic limit is given by (\cite{woynarovich91,essler-book,oelkers06} and references therein)
\begin{equation}\label{energyBA}
E=-\int_{-Q}^Q 2 \cos k \rho_1(k)-\int_{-B}^B 4 \textrm{Re} \sqrt{1-(\lambda-i u)^2}\rho_2(\lambda),
\end{equation}
where $\rho_1(k)$ and $\rho_2(k)$ are spectral functions satisfying the following coupled integral equations:
\begin{eqnarray}
&\rho_1 (k)&=\frac{1}{2 \pi} -\frac{1}{2 \pi} \cos k  \int_{-B}^B 
K_1(\sin k- \lambda^\prime)\rho_2(\lambda^\prime)d\lambda^\prime \nonumber  \\
&\rho_2 (\lambda)&=\frac{1}{\pi}   \textrm{Re} \frac{1}{\sqrt{1-(\lambda-i u)^2}} \nonumber\\
&&-\frac{1}{2 \pi} \int_{-Q}^Q K_1(\lambda-\sin k^\prime)\rho_1(k^\prime)dk^\prime  \nonumber\\
&&-\frac{1}{2 \pi} \int_{-B}^B K_2(\lambda-\lambda^\prime)\rho_2(\lambda^\prime) d\lambda^\prime. 
\label{spectralBA}
\end{eqnarray}
Here $K_\nu(x)=2 \nu u/(\nu^2 u^2+x^2)$, with $u=|U|/4$.
The integration limits $B,Q$ are related to the total density $n=n_\uparrow + n_\downarrow$ and the density difference $s=n_\uparrow - n_\downarrow$ by the conditions
\begin{equation}\label{densitiesBA}
\int_{-Q}^Q \rho_1(k) dk= s,   \;\; \;\;\;\;  2 \int_{-B}^B \rho_2(\lambda) d\lambda= n-s. 
\end{equation}

Equations (\ref{energyBA})-(\ref{densitiesBA}) are only valid below unit filling, $n \leq 1$. For
higher concentrations the ground state energy can be obtained through a particle-hole transformation $h_\sigma=c_\sigma^\dagger$ and is given by $E(n,s)=E(2-n,s)+U(n-1)$. From the energies,  we calculate the
averaged chemical potential and the effective magnetic field: 
\begin{equation}\label{change}
\mu=\frac{\partial E}{\partial n},\;\;\;\;\; h= \frac{\partial E}{\partial s}.
\end{equation}
Interpreting Eq.~(\ref{change}) as a change of variables, one obtains the phase diagram  
of the Hamiltonian ~\eqref{one}, as shown in Fig.~\ref{fig0} for a specific value of the interaction $U=-5t$ (compare \cite{essler-book}). 
We can distinguish different phases for a low filling $n=N/L\leq1$: (i) the vacuum (V), corresponding to $s=n=0$, (ii) the unpolarized phase (ED) characterized by equal densities of majority and minority fermions, 
(iii) the partially polarized phase (PP), corresponding to $0<s<\textrm{max}(n,2-n)$; (iv) the fully polarized phase with $n<1$ (FP$_1$), and (v) the fully polarized phase with $n=1$ (FP$_2$). 
The remaining phases at higher densities correspond to situations where at least one of the component forms a band insulator.

\begin{figure}[t]
\epsfig {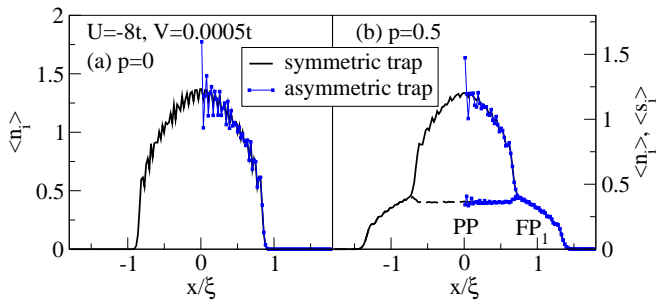}
\caption{
(color online)
Particle density $\langle n_i\rangle $ (solid lines) and local polarization $\langle s_i\rangle $ (dashed lines)
for $U=-8t$, $V=0.0005t$, and (i) a symmetric trap with $N=80$ (lines)
and (ii) an asymmetric trap with $N=40$ (lines with symbols). (a) $p=0$; (b) $p=1/2$.
}
\label{fig1}
\end{figure}

\subsubsection{Trapped system}

 The density profiles of the gas in a shallow trap can be calculated via the local density approximation, assuming that the system is locally homogeneous: 
\begin{eqnarray}\label{lda}
\mu \lbrack n(x),s(x)  \rbrack &=&  \mu^0 -V x^2  \label{eq:lda1}\\
h \lbrack n(x),  s(x) \rbrack &=& h^0,         \label{eq:lda2}
\end{eqnarray}
where $\mu^0$ and $h^0$ are the chemical potential and the magnetic field at the trap center. In the second line of
Eq.~(\ref{lda}), we have exploited that the  external potential is the same for both components, corresponding to vertical trajectories in the phase diagram of Fig.~\ref{fig0}. 
The constants $\mu^0,h^0$ are fixed by the overall spin populations $N_\sigma$ through the normalization conditions
\begin{equation}\label{norm}
N=\int n(x)dx, \;\;\;\; N_\uparrow-N_\downarrow=\int s(x)dx.
\end{equation}
By inverting Eq.~(\ref{change}) we see that the normalization conditions (\ref{norm}) can be written as 
$N=\int n[\mu^0-V x^2,h^0] dx $, with an analogous expression for the density difference. Introducing the new coordinate 
$y=\sqrt{V} x$, we get $N\sqrt{V}=\int n[\mu_0-y,h^0]$, showing that the constants $\mu^0,h^0$
(and through them, the density profiles) are completely determined by three parameters: the effective density $\rho_{\mathrm{eff}}=N\sqrt{V}$, 
the spin polarization $p=(N_\uparrow-N_\downarrow)/N$, and the interaction strength $U$.

In this article we mainly consider configurations where the core of the cloud is partially polarized, corresponding to trajectories starting from the PP phase.
Then, LDA predicts the existence of two shell structures: a partially polarized (PP) core with either (i) fully polarized (FP$_1$) or (ii) fully paired (ED) wings
(see the vertical lines in Fig.~\ref{fig0} for an illustration). 
The dotted line in Fig.~\ref{fig0} corresponds to the special case  $h=\sqrt{U^2+16}/2-2$, where the PP phase extends to the edge of the cloud, in analogy with the continuum case \cite{orso07,hu07}. 
This yields a critical value of the spin polarization $p=p_c$, 
separating the regime with  fully paired $(p<p_c)$ from the one with fully polarized wings $(p>p_c)$.
At larger densities, more shell structures are possible, e.g.,
a band insulator of the majority spins in the center with two surrounding shells \cite{feiguin07}.

\begin{figure}[t]
\epsfig {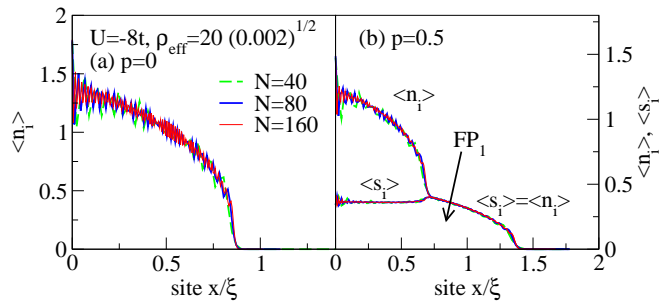}
\caption{
(color online)
Density and polarization profiles in the asymmetric trap for different particle numbers $N=40,80,160$ at a fixed
effective density $\rho_{\mathrm{eff}}=N\sqrt{V}$ at $U=-8t$ (we choose $V$ such that at each $N$, $\rho_{\mathrm{eff}}=N\,\sqrt{V}=20\sqrt{0.002}=$const).
(a) $p=0$, (b) $p=0.5$. For the largest $N$ and $p=1/2$, we use chains with $L=300$ sites.}
\label{fig2}
\end{figure}

\subsection{DMRG simulations} 
The density profiles of the gas in the trap can be calculated to a high accuracy using the DMRG algorithm.
We use a standard implementation for ground state calculations \cite{schollwoeck05} and up to $600$ states.
Occasionally, it was necessary to perform an unusually large number of  sweeps (sometimes more than 20) to obtain
converged results. For a  Hamiltonian with site-dependent parameters,  it is convenient to grow the system in a linear fashion, say from the left to the right,
during the warm-up part of the DMRG algorithm. The largest system size used in this work is $L=300$ sites for $p=0.5$, $U=-8t$ and $N=160$ particles. In that case, typical 
discarded weights in the center of the system are smaller than $10^{-10}$. 
 In the two following subsection we provide details on the set-up used in the
DMRG runs, such as the choice for the trapping potential and a finite-size scaling analysis.

\subsubsection{Asymmetric vs symmetric traps} 

We here consider two types  of a harmonic confining potential parameterized by a constant $V$,
namely one that is symmetrically placed in the optical lattice [see Eq.~(\ref{eq:strap})] and one that 
traps particles in the left part of the chain:
\begin{equation}
H_{\mathrm{trap}} =  V\sum\limits^L_{i=1}x^2n_i\,. \label{eq:atrap}
\end{equation}
We will next show that for the purpose of studying the behavior in the wings, the symmetric
Eq.~\eqref{eq:strap} and the asymmetric set-up Eq.~\eqref{eq:atrap} yield the same quantitative behavior. 
Note that in the figures, we shall display the results for the symmetric trap  shifted by $L/2$ lattice sites to the left with respect to Eq.~(\ref{eq:strap}).
The advantage of using  the asymmetric trap is that  it allows us 
to reach much lower densities and larger particle numbers without resorting to large system sizes, and at less computational costs.
Note that for the same trapping amplitude $V$, the particle numbers  in the symmetric trap 
Eq.~\eqref{eq:strap} are twice as large as the ones  in the corresponding asymmetric set-up.

Figure~\ref{fig1} shows the comparison between the symmetric and the asymmetric trap for the parameters
of  Ref.~\cite{feiguin07} ($U=-8t$, $V=0.002t$, $N=40$ in the case of the symmetric trap) for the 
balanced case [$p=0$, panel (a)] and the imbalanced case [$p=1/2$, panel (b)]. The open boundary that is present when
we use the asymmetric set-up clearly induces strong additional oscillations in the density profile, both at $p=0$
and $p=0.5$. The main point is, however, that away from the left boundary and towards the edge of the cloud,
the density and polarization profiles computed from either the symmetric or the asymmetric trap coincide. 
In the case of $p=0.5$, the system enters the fully polarized wings at the same distance from the trap center,
independently of the particle number.
This justifies the use of the asymmetric trap in our analysis of the shell structure, which we shall 
use in the remainder of this work.

\subsubsection{Scaling analysis in the particle number}

As a next step we study the finite-size scaling of the density and polarization profiles. More precisely, 
we increase the particle number but keep the effective density $\rho_{\mathrm{eff}}=N\sqrt{V}$ fixed \cite{rigol04c,rigol04d} (the effective density is given for the {\it asymmetric} trap unless stated otherwise).
Site-dependent quantities should then be plotted versus $x/\xi$, where $\xi=(V/t)^{-1/2}$ is the characteristic length scale for a given
effective density. The results of such an analysis are presented in Fig.~\ref{fig2} for the parameters  of Fig.~\ref{fig1},
with particle numbers $N=40,80,160$ in the asymmetric trap.
As expected, all curves computed for different particle numbers fall on top of each other.
We further see   that the amplitude of
the boundary-induced  oscillations becomes much weaker the larger the polarization or the larger $N$ is 
 and therefore, the presence of the oscillations does not pose a problem for our analysis.

\begin{figure}[t]
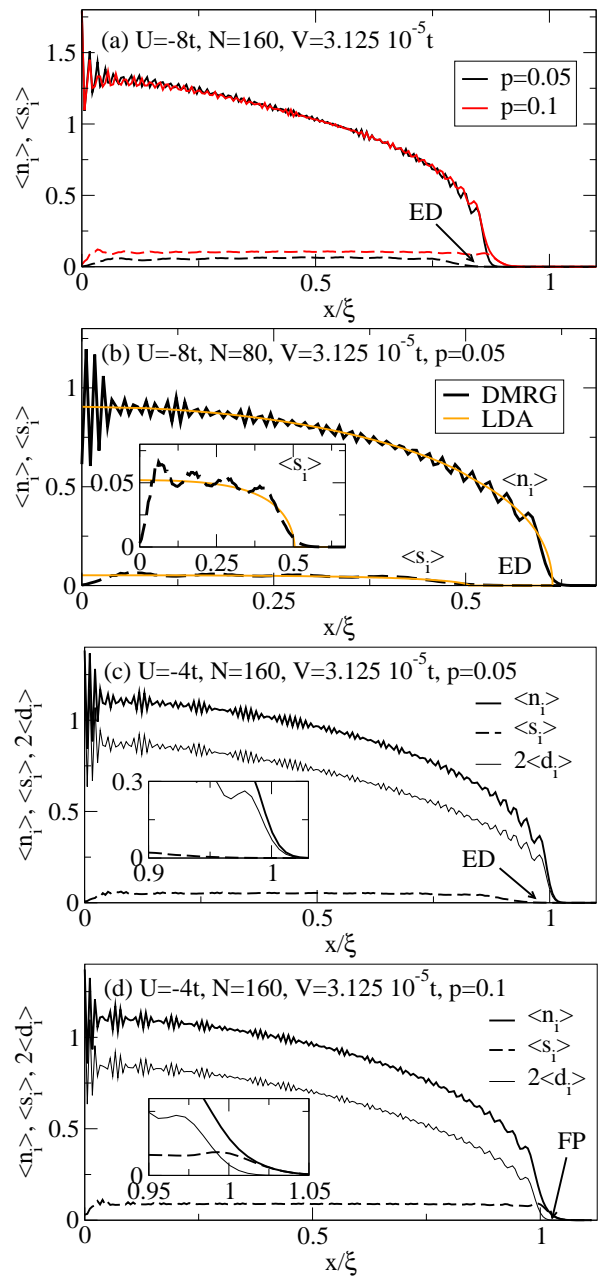

\epsfig {file= figure4a.eps,width=0.9\columnwidth}
\epsfig {file= figure4b.eps,width=0.9\columnwidth}
\epsfig {file= figure4c.eps,width=0.9\columnwidth}
\epsfig {file= figure4d.eps,width=0.9\columnwidth}
\caption{
(color online)
Shell structure in an asymmetric trap for (a) $U=-8t$, $ N=160$, (b) $U=-8t$, $N=80$
and (c), (d) $U=-4t$, $N=160$. In all cases, $V=3.125\, 10^{-5}t$. Thick solid lines: density profile $\langle n_i\rangle $, 
dashed lines: polarization profile $p_i$, thin solid lines in (c), (d): double occupancy $2\langle d_i \rangle $.
Panels (a), (c), and (d) show DMRG results only while (b) includes LDA results (thin lines) for comparison.
The inset in (b) shows the density difference and the insets in (c), (d) give an enlarged view of the interface regions.
}
\label{fig3}
\end{figure}

\section{Results}
\label{sec:results}
\subsection{Fully paired wings} 

In this section we will focus on $N=160$ and $N=80$ particles, with a trapping potential
 of $V=3.125 \,10^{-5}t$. The change from fully paired (ED, equal densities) wings to fully polarized (FP$_1$) wings is expected to occur
at small polarizations, and we hence concentrate on $p\lesssim 0.2$. Figure~\ref{fig3}(a) 
contains the results for $\rho_{\mathrm{eff}}=20\,\sqrt{0.002}$, i.e., the original density studied in  Ref.~\cite{feiguin07}, 
but with a much larger number of particles.
While at $p=0.1$, we still clearly observe FP wings, at $p=0.05$, the local polarization practically 
vanishes at the edge of the cloud. Therefore, for these parameters, the critical polarization for the change in the shell
structure must be $0.05<p_c< 0.1$, and further analysis shows that it is $p_c\approx 0.07$. 
Therefore, if there are too few particles in the trap, say $N=40$, then the fully paired wings are very difficult to observe
since such a $p_c$ would correspond to one of the smallest possible polarizations. Moreover, effects due to a finite particle number become more relevant
for small $N$ and/or tighter traps, and LDA does not necessarily  quantitatively apply in this regime (see also Ref.~\cite{casula08}). In fact, numerically
exact DMRG calculations do not yield  evidence for fully paired wings at, for instance,  $N=40$ and $V=0.005t$ \cite{feiguin07}.

By going to half as many particles yet while keeping the trapping amplitude constant at $V=3.125\, 10^{-5}t$ 
and thus decreasing the effective density, the fully paired wings become better visible, as we demonstrate in Fig.~\ref{fig3}(b) for $p=0.05$.
There, the local polarization vanishes at a distance from the center of the trap close to $x/\xi\sim 0.5$, whereas the cloud extends up to $x/\xi\sim 0.6$. Clearly,  the volume of the fully paired wings increases going to a lower effective density.

Another way of influencing  the volume of the fully paired wings is by going to smaller values of the interactions $|U|$. The respective results
for $U=-4t$ and the  same density and particle number as in  Fig.~\ref{fig3}(a) are shown in Figs.~\ref{fig3}(c) and (d) for
$p=0.05$ and $p=0.1$, respectively. In this example,
 we find evidence for the fully paired wings at $p=0.05$, while at $p=0.1$, we see fully polarized wings.
The fully paired wings show up as first, a region of practically vanishing local polarization and second, the double occupancy
$\langle d_i\rangle =\langle n_{i\uparrow} n_{i\downarrow} \rangle $ [thin solid lines in Figs.~\ref{fig3}(c) and (d)]  remains finite there. By contrast, whenever
 we enter the fully polarized wings, the double occupancy goes to zero.
Summarizing, so far, we have illustrated that the fully paired wings indeed show up in 1D optical lattices at small 
polarizations  in shallow traps, similar to the case of the associated continuum model \cite{orso07,hu07,casula08}.
Decreasing the effective density  stabilizes this shell structure in the sense that, the relative volume
of the fully paired wings increases [compare Figs.~\ref{fig3}(a) and (b)].

\begin{figure}[t]
\epsfig {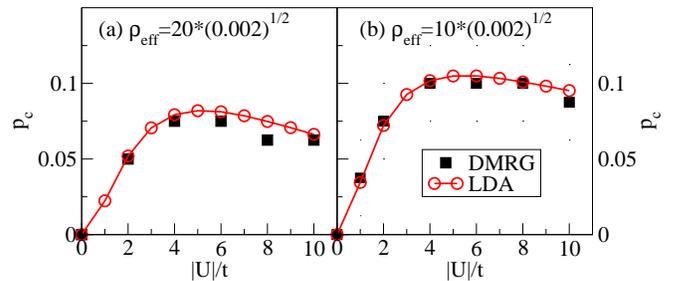}
\caption{
(color online)
Critical polarization $p_c$ for the change in the shell structure from fully paired wings at $p<p_c$ 
to fully polarized wings at $p>p_c$, vs. interaction $|U/t|$  for two different effective densities  
$\rho_{\mathrm{eff}}=N\, \sqrt{V}$ ($V=3.125\, 10^{-5}t$) with (a) $N=160$ and (b) $N=80.$ 
 The DMRG (squares) results were extracted from runs with an asymmetric trap.
 LDA results are displayed with circles. The non monotonic behavior of the critical polarization
is a unique feature of the lattice, not found for the continuum model.}
\label{fig5}
\end{figure}

Next, we compare the density profiles calculated by DMRG and by LDA, where the latter makes use of the exact phase diagram for the 1D Hubbard model (Fig.~\ref{fig0}, see \cite{essler-book}). This is shown in Fig.~\ref{fig3}(b) for $U=-8t$ and a polarization of $p=0.05$ at which the wings are fully paired. The agreement between LDA and DMRG is very convincing which is expected in the case of a shallow trap as considered here ($V=3.125 \,10^{-5}t$). Most importantly for us here, 
both methods show the fully paired wings and the cloud's spatial extension for both the density and density difference agree quite well.
From the figure, some known shortcomings of LDA are obvious:
first,  LDA fails to reproduce the correct decrease of the density profile  at the edge of the cloud. 
This implies that within LDA it is possible to simply define the radius of the cloud by its extension, whereas for a finite particle number this quantity fluctuates.
Second, the incommensurate oscillations in the
density induced by the lattice are not captured by LDA either \cite{molina07}. While these discrepancies are  not crucially relevant to the analysis
carried out here, there is another feature that is absent in the LDA, namely the $2Q_{\mathrm{FFLO}}$ oscillation in the density difference. This is obvious since
LDA assumes that the gas is locally homogeneous  whereas in DMRG calculations and in experiments, one deals with a finite particle number.

The $2Q_{\mathrm{FFLO}}$ modulation is
due to the fact that majority fermions reside in the nodes of the quasi-condensate \cite{mizushima05}, which has previously been demonstrated  with DMRG \cite{feiguin07,tezuka08,hm09c}. In other words, 
the incommensurate 1D FFLO state is accompanied with a spin-density wave (see Ref.~\cite{roscilde09}), as is illustrated in the inset of Fig.~\ref{fig3}(b). This feature, which is visible in a trap, where
the quasi-condensate is pinned,  could in principle be observed in
experiments with trapped two-component Fermi gases \cite{mizushima05,hm09c}. This highlights the importance of using exact numerical techniques such as DMRG and QMC that can take the
trap into account exactly to complement analytical methods. Nevertheless, LDA combined with the {\it exact} solution of the phase diagram \cite{orso07,hu07} has been very successful in 
 predicting the shell structures in one dimension, which for the continuum case has been observed experimentally \cite{liao09}.

\subsection{Critical polarization}

We now present  our theoretical predictions for the critical polarization $p_c$ associated to the
change in the shell structure of the density profiles, from fully paired $(p<p_c)$
to fully polarized $(p>p_c)$ wings. For a shallow trap, this quantity depends only on the 
values of the interaction $U$ and the effective density $\rho_\textrm{eff}=N\sqrt{V}$.

In Fig.~\ref{fig5} we plot the critical polarization as a function of $|U|$ comparing DMRG (squares) with LDA (circles), for $V=3.125 \, 10^{-5}t$  and two different particle numbers $N=160$ and $N=80$.
From the DMRG simulations, we estimate $p_c$ as the point at which, by visual inspection, we can clearly resolve
regions in the wings with $\langle s_i \rangle= \langle n_i\rangle$. Due to the finite particle number, these 
results carry uncertainties of $\delta p_c \approx \pm 0.0125,0.03$ for $N=80, 160$, respectively.

 The agreement between the  DMRG results and LDA is obviously very good. For $|U| \lesssim 5t$, the critical polarization is an monotonically increasing function of the interaction, in analogy with the continuum model \cite{hu07,orso07}. However, for stronger interactions the
critical polarization in a lattice reaches a maximum value and then decreases at large $|U|$, in contrast with the continuum case \cite{hu07,orso07}.

The physical explanation  is that in this regime the kinetic energy of the pairs is suppressed, since their effective mass becomes increasingly large $(\sim|U|/t^2)$ compared to that of single fermions $(1/t)$. As a consequence, the
energy needed to break a pair is given by $|U|$, independently of the particle density. 
This means that 
the boundary between the fully paired (ED) and the partially polarized phase in Fig.~\ref{fig0} (the line connecting the points A,B, and C counterclockwise)
 reduces to the dotted vertical line for $U\to -\infty$, implying the absence of fully paired wings in a trap.

Figure~\ref{fig5} further suggests that for the chosen parameters, the maximum critical polarizations are $p_c^{\mathrm{max}}\approx 0.08$ and $p_c^{\mathrm{max}} \approx 0.1$, respectively.
These values are  smaller than the largest critical polarization $p_c^{\mathrm{max}} \approx 0.2$
found in the continuum model.

\begin{figure}[t]
\epsfig {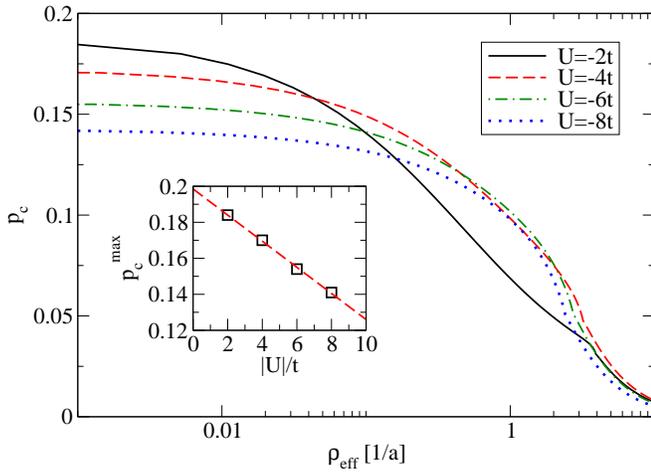}
\caption{
(color online)
Critical polarization $p_c$ for the change in the shell structure from fully paired wings at $p<p_c$ 
to fully polarized wings at $p>p_c$ as a function of effective density $\rho_{\mathrm{eff}}=N \sqrt{V}$ (given in units of the inverse lattice spacing $1/a$). We display LDA results
for several $U$ as indicated in the legend. Inset: maximum $p_c^{\mathrm{max}}=\lim_{\rho_{\mathrm{eff}}\to 0 }p_c$ vs $|U|$. The dashed line is a linear fit to the LDA data.}
\label{fig6}
\end{figure}

In Fig.~\ref{fig6} we plot LDA results for the critical polarization versus the 
effective density for different values of the interaction. We see  that the polarization increases by decreasing the effective density and saturates at a maximum value of   $p_c^{\mathrm{max}}(U)$. Then, if one sends $U$ to zero as well, this maximum will approach $p_c^{\mathrm{max}}(U)\to 0.
2$ as is illustrated in the inset of Fig.~\ref{fig6}, consistent with the findings reported for the continuum limit \cite{hu07,orso07}. 

\section{Summary and Discussion}
In summary, we  investigated the density profiles of trapped spin-imbalanced    Fermi gases in a 1D optical lattice. We demonstrated that, when performed on sufficiently large systems,  numerically exact DMRG
simulations are fully consistent with  LDA results. 
 In particular, we reported numerical evidence for the fully paired wings that were predicted to exist at small polarizations \cite{hu07,orso07}. However, differently from the continuum case, we found   
that  the upper critical polarization $p_c$ for the appearance of the fully paired wings is a non monotonic function of the interaction strength $U$ at a fixed filling. This behavior is a direct consequence of the increased inertia of the pairs
in the strong coupling regime $(|U|\gg 4t)$.    


\acknowledgments

We thank David Huse and Xiwen  Guan for fruitful discussions.

\end{document}